\documentclass{iau}
\usepackage{graphicx}

\title[Progenitors of Magnetars and Hyperaccreting Magnetized Disks] 
{Progenitors of Magnetars and Hyperaccreting Magnetized Disks}

\author[Yi Xie \& Shuang-Nan Zhang]   
{Yi Xie$^1$
 \and  Shuang-Nan Zhang$^{1,2}$}

\affiliation{$^1$National Astronomical Observatories, Chinese Academy Of Sciences,\\ Beijing, 100012, P. R.China
\\ email: {\tt sourcexieyi@gmail.com } \\[\affilskip]
$^2$Key Laboratory of Particle Astrophysics, Institute of High
Energy Physics, Chinese Academy of Sciences, \\ Beijing, 100012, P.
R. China \\email: {\tt zhangsn@ihep.ac.cn}}

\pubyear{2012}
\volume{290}  
\jname{Feeding compact objects: Accretion on all scales}
\editors{C.M. Zhang, T. Belloni, M. M\'endez \& S.N. Zhang, eds.}

\begin{document}

\maketitle

\begin{abstract}
We propose that a magnetar could be formed during the core collapse of massive stars or coalescence of two
normal neutron stars, through collecting and inheriting the magnetic fields magnified by hyperaccreting disk.
After the magnetar is born, its dipole magnetic fields in turn have a major influence on the following
accretion. The decay of its toroidal field can fuel the persistent X-ray luminosity of either an SGR or AXP;
however the decay of only the poloidal field is insufficient to do so.

\end{abstract}

\firstsection 
\section{Introduction}
Neutron stars (NSs) with magnetic field $B$ stronger than the quantum critical value, $B_{\rm cr}=m^2
c^3/e\hbar\approx 4.4\times 10^{13}~$G, are called magnetars. Observationally, a magnetar may appear as a soft
gamma-ray repeater (SGR) or an anomalous X-ray pulsar (AXP). It is believed that its persistent X-ray luminosity
is powered by the consumption of their $B$-decay energy (e.g. \cite[Duncan \& Thompson 1992]{DuncanThompson92};
\cite[Paczynski 1992]{Paczynski92}). However, the formation of strong $B$ of a magnetar remains unresolved, and
the possible explanations are divided into three classes: (i) $B$ is generated by a convective dynamo
(\cite[Duncan \& Thompson 1992]{DuncanThompson92}); (ii) $B$ is essentially of fossil origin (e.g.
\cite[Ruderman 1972]{Ruderman72}; \cite[Ferrario \& Wickramasinghe 2006]{FerrarioWickramasinghe06}); and (iii)
$B$ has evolved from some radio pulsars after many glitches (e.g. \cite[Lin \& Zhang 2004]{LinZhang04}). In this
paper we propose that a magnetar is likely formed during the hyperaccreting process of NSs.
\section{The Formation of a Magnetar}
Stage 1: \emph{$B$ magnified significantly in hyperaccreting process.} It is generally believed that $B$, which
could be magnified by magneto-rational instability (\cite[Balbus \& Hawley 1991]{BalbusHawley91}) or dynamo
process on the disk, affects the angular momentum transfer effectively via a variety of modes (e.g.
\cite[Blandford 1976]{Blandford76}; \cite[Blandford \& Payne 1982]{BlandfordPayne82}; \cite[Balbus \& Hawley
1991]{BalbusHawley91}). For hyperaccretion disks, $B$ must be magnified extremely strongly so that the accretion
rate can be as high as $\dot M \sim 0.1~\rm M_\odot s^{-1}$. In detailed magnetohydrodynamic simulations of
binary NS merger processes, $B$ is amplified by Kelvin-Helmholtz instabilities, and finally grows to
$2\times10^{15}~$G (\cite[Price \& Rosswog 2006]{PriceRosswog06}; \cite[Giacomazzo, Rezzolla \& Baiotti
2009]{GiacomazzoRezzollaBaiotti09}). The basic equations describing the magnetized hyperaccretion flow consist
of the angular momentum equation, energy equation and the equation of state (Di Matteo et al. 2002; LoveLace,
1995; Xie et al. 2009).

Stage 2: \emph{$B$ inherited by the central NS.} The magnified $B$ in the form of loops and chaotic
configuration is sheared by the disk differential rotation, leading to open field lines (\cite[Romanova
1998]{Romanova98}). The fields froze-in to highly conducting disk plasma, are transported to the central object
by the inward motion of the disk matter (e.g. \cite[Ghosh \& Abramowicz 1997]{GhoshAbramowicz97}; \cite[Spruit
\& Uzdensky 2005]{SpruitUzdensky05}). We get the magnetic flux of the whole disk by $\Phi=\int^{R_{\rm
out}}_{R_{\rm in}}2\pi r(B_{z})dr\simeq 1.4\times 10^{29}~{\rm cm}^{2}{\rm G}.$ If the poloidal field $B_{z}$
anchored at the disk could be mainly collected and inherited by the central NS, the field strength near its pole
area can be as high as $B_\ast\sim\Phi/2\pi r_{\ast}^2\simeq 1.5\times 10^{16}~\rm G$. This means that a
magnetar is formed during the hyperaccreting process. For the new-born magnetar, its dipole fields are mainly
composed of $B_{z}$ and radial component $B_{r}$ of the disk, since they are closely related to the toroidal
current in the disk, and the toroidal component $B_{\phi}$ corresponds to multipolar fields. The ratios of the
field components of the disk are quite similar to the field configuration of a magnetar.

Stage 3: \emph{$B$-Decay and Dominant Toroidal Field Component.} Several avenues exist for $B$-decay in isolated
NSs: ohmic decay, ambipolar diffusion, and Hall drift (\cite[Goldreich \& Reisenegger
1992]{GoldreichReisenegger92}; \cite[Heyl \& Kulkarni 1998]{HeylKulkarni98}). Depending on the strength of $B$,
each process may dominate the evolution. It has been found that the power released by the decay of dipole fields
cannot fuel the persistent X-ray luminosity of several magnetars. However, the power released by the decay of
the toroidal component is sufficient for persistent X-ray luminosity of all magnetars. This could be an indirect
observational evidence for the existence of their dominant toroidal $B$.
\section{Conclusions}
We suggest that a magnetar could be formed during the coalescence of binary NSs or collapse of a massive star,
through collecting and inheriting the magnetic fields magnified by the hyperaccreting disk. After the magnetar
is born, its dipole magnetic fields in turn have a major influence on the following accretion. The decay of
toroidal fields can fuel the persistent X-ray luminosity of either an SGR or AXP; however the decay of only the
poloidal field is insufficient to do so.

\end{document}